\documentclass[twocolumn,prb]{revtex4}

\usepackage{graphicx}
\usepackage{bm}

\begin{document}

\title{Extracting d-orbital occupancy from
magnetic Compton scattering   \\
in bilayer manganites}

\author{B. Barbiellini$^1$, P. E. Mijnarends$^{1,2}$,
S. Kaprzyk$^{1,3}$, A. Bansil$^1$,
Yinwan Li$^{4,5}$, J.F. Mitchell$^5$ and P. A. Montano$^{4,6}$
}
\affiliation{
$^1$Physics Department, Northeastern University, Boston MA 02115 \\
$^2$ Department of Radiation, Radionuclides \& Reactors,
Faculty of Applied Sciences, \\
Delft University of Technology,
Delft, The Netherlands \\
$^3$Academy of Mining and Metallurgy AGH, 30059 Krak\'ow, Poland\\
$^4$Department of Physics, University of Illinois, Chicago IL 60680\\
$^5$Materials Science Division, Argonne National Laboratory, Argonne IL
60439\\
$^6$Scientific User Facilities Division, U.S. Department of Energy, 1000
Independence Avenue, Washington DC 20585-1290\\
}

\begin{abstract}

We consider the shape of the magnetic Compton profile (MCP),
$J_{mag}(p_z)$,
in La$_{1.2}$Sr$_{1.8}$Mn$_2$O$_7$ for momentum transfer $p_z$
along the [110] direction and the
associated reciprocal form factor $B(r)$ defined by taking the
one-dimensional Fourier transform of $J_{mag}(p_z)$. $B(r)$
is shown to contain a
prominent dip at $r\approx$ 1 $\mbox{\AA}$, where the minimum value
$B_{min}$ of $B(r)$
can be related to the occupancies of the $e_g$ orbitals of
$d_{x^2-y^2}$ and $d_{3z^2-r^2}$
symmetry in the system. We illustrate our procedure
in detail by analyzing
the measured MCP at 5K and the MCP computed within the framework of the
local spin density approximation (LSDA) and comment on the differences
between the measured and computed $e_g$ occupancies as a reflection of the
limitations of the LSDA in treating electron correlation effects.

\end{abstract}

\maketitle

\section{Introduction}

The double-layered manganites La$_{2-2x}$Sr$_{1+2x}$Mn$_2$O$_7$ have drawn
a great deal of attention as model systems that display a wide range of
transport and magnetic properties and undergo a variety of phase
transitions as a function of temperature, doping and magnetic field. In
La$_{1.2}$Sr$_{1.8}$Mn$_2$O$_7$ (i.e., $x = 0.40$), investigated in this
study, the CMR effect is a factor of $\sim 200$ at 129 K (just above the
Curie temperature $T_c \sim 120$ K) under a magnetic field of 7 T, and
even at low fields the resistance changes by $\sim 200
\%$.\cite{moritomo96} Neutron studies\cite{hirota98} in this compound
show that below $T_c$ the spins are aligned ferromagnetically within the
MnO planes, but are canted antiferromagnetically between adjacent Mn-O
planes.

Compton scattering refers to inelastic x-ray scattering in the deeply
inelastic regime and it is well known that this technique directly probes
the ground state momentum density $\rho({\mathbf p})$ of the many body
electronic system through a measurement of the so-called Compton profile,
$J(p_z)$, where $p_z$ is the momentum transferred in the scattering
process.\cite{cp_book,kaplan03} 
X-rays interact not only with the electron charge but
also with the electron spin. Therefore, the majority (minority) spin
Compton profile $J_{\uparrow}$ ($J_{\downarrow}$) can be extracted by 
measuring
two Compton scattering cross sections: one with the direction of the
magnetization of the specimen parallel to the scattering vector,
and the other with the magnetization anti-parallel.
Thus, in ferro- or ferrimagnetic materials one can zoom
in on the magnetic properties via the magnetic Compton profile (MCP)
$J_{mag}({\bf p})$ obtained by taking the difference between spin up
(majority) and spin down (minority) Compton profiles
\begin{equation}
J_{mag}(p_z) =
            J_{\uparrow}(p_z) - J_{\downarrow}(p_z)~.
\label{eq1}
\end{equation}
$J_{mag}$ can be expressed in terms of a double integral of the
spin density, $\rho_{mag}({\mathbf p})$ as
\begin{equation}
J_{mag}(p_z)= \int \int  \rho_{mag}({\mathbf p}) dp_x dp_y,
\label{eq2}
\end{equation}
where
$\rho_{mag}({\mathbf p})\equiv \rho_{\uparrow}({\mathbf p})
- \rho_{\downarrow}({\mathbf p})$.
The possibility of using magnetic Compton scattering (MCS) to determine
the momentum distribution of magnetic electrons was recognized quite
early,\cite{platzman65,sakai76} but since the scattering cross-section in
the magnetic channel is typically several orders of magnitude smaller than
for charge scattering, MCS experiments have become practical only in the
last few years with the availability of high-energy, circularly polarized,
X-rays at the synchrotron light sources.\cite{montano95}

In a recent MCS study of La$_{1.2}$Sr$_{1.8}$Mn$_2$O$_7$, we have focused
on the [110] MCP and shown that the shape of the [110] MCP contains a
remarkable signature of the occupancy of the $d_{x^2-y^2}$
electrons.\cite{li04} By using a high magnetic field of 7T to maintain an
electronically homogeneous ferromagnetic phase, we demonstrated that
variations in the occupancy of the $d_{x^2-y^2}$ orbitals can be adduced
from the [110] MCPs measured at different temperatures. The purpose of the
present article is to expand on the discussion
of Ref.~\onlinecite{li04}  concerning how
the shape of the [110] MCP can be analyzed in terms of the reciprocal form
factor $B(r)$, which is related to the MCP via a one-dimensional Fourier
transform along the direction of the scattering vector. In this
connection, detailed forms of the relevant $B(r)$
functions for various $t_{2g}$
and $e_g$ orbitals are given and their characteristic behavior with $r$ is
delineated. Since our purpose is to illustrate the fitting procedure for
interpreting the shape of the $B(r)$, we only discuss the experimental
$B(r)$
at 5K and the corresponding theoretical curve obtained using the local
spin density approximation (LSDA). We also comment on the discrepancies
between these computed and measured $B(r)$'s as a reflection of the
shortcomings of the LSDA in treating electron correlation effects.

Concerning other MCP studies of the La-manganite, Koizumi {\em et
al.}\cite{koizumi01,koizumi03} have
investigated the doping dependence of the [100]
and [001] MCPs under a relatively low magnetic field of
2.5 T. They analyze their MCPs in terms of atomic and cluster type
computations of the momentum density to gain insight into the occupation
of magnetic orbitals in the system.  All existing MCP data
on the double-layered manganite
have been taken at a momentum resolution of around 0.4 a.u. Recently,
Mijnarends {\em et al.} \cite{mijna05} have investigated the MCPs in
La-manganite along the three high symmetry directions and have shown that
the
resolution of about 0.4 a.u. is not sufficient for investigating Fermi
surface signatures in the MCPs.

This article is organized as follows. Section \ref{sec:meth} gives some
technical details of our LSDA-based MCP computations and
briefly introduces the $B(r)$ function which is the basis of our
analysis. Section~\ref{sec:res} presents and discusses the details of how
we analyze and fit the $B(r)$ function.  Section \ref{sec:sum} makes a few
concluding remarks. As already indicated, in this article we only discuss
the 5 K measurements taken under a magnetic field of at 7T; for technical
details of the experimental measurements, the reader is referred to Refs.
\onlinecite{montano95} and \onlinecite{li04}.

\section{Computational Details}
\label{sec:meth}

In order to obtain momentum densities and Compton profiles, the electronic
structure of LaSr$_2$Mn$_2$O$_7$ was first obtained within an all-electron
charge and spin self-consistent KKR framework. The reader is referred to
Refs. \onlinecite{kaprzyk90,bansil91,bansil99,footnote0.5} for
details of our band structure computations. The formalism for computing
momentum densities is discussed in Refs.
\onlinecite{mijnarends76,mijnarends79,mijnarends90,bansil81,mijnarends95}.
The lattice data were taken from Seshadri {\em et al.}\cite{seshadri97}
Exchange-correlation effects were incorporated
within the local spin density approximation (LSDA) \cite{lsda}.
Doping effects were treated within
the rigid band approximation where the band structure of 
La$_{2-2x}$Sr$_{1+2x}$Mn$_2$O$_7$ is assumed to be the same as that of 
LaSr$_2$Mn$_2$O$_7$. More specifically, we have modeled the doped
system by adjusting 
the Fermi energy $E_F$ in the majority spin band to accommodate
the proper number of electrons for the $x=0.4$ case.\cite{huang00}

A prominent feature of the band structure is the
presence of a large exchange splitting on Mn, which yields a nearly
unoccupied minority spin Mn-3d band.\cite{mijna05,huang00,deboer99} An
analysis of the density of states shows that the $t_{2g}$ states lie at a
binding energy of about 1 eV, while the $e_{g}$ states are placed
essentially at the Fermi energy. Therefore our {\em ab-initio} calculation
is consistent with molecular models which give a splitting of about 1 eV
between the $t_{2g}$ and $e_{g}$ levels.\cite{owen66}
We should keep in mind however
that in a band structure scheme the $t_{2g}$ and $e_{g}$ characters of
Bloch wavefunctions vary continuously across various bands and that there
is no strict gap between these levels for either the majority or the
minority bands.

If we assume that the $3.6$ Mn electrons effectively retain their atomic
character, then we would expect that these electrons will occupy the
up-spin states following Hund's first rule. In particular, the first three
electrons will go into the three lower lying up-spin $t_{2g}$ orbitals
$(d_{xy}, d_{xz}, d_{yz})$, while the remaining 0.6 electrons will occupy
the $e_g$ orbitals ($d_{x^2-y^2}$, $d_{3z^2-r^2}$). Moreover,
since the $t_{2g}$
and $e_g$ states are separated by a large energy gap of about $1$ eV, which
is much larger than $kT$, the occupancy of $t_{2g}$ states will change
little
with $T$, and the effect of temperature will mainly be to redistribute the
$0.6$ $e_g$ electrons between the $d_{x^2-y^2}$ and $d_{3z^2-r^2}$
orbitals. A more quantitative handle on the occupancy of magnetic orbitals
can be obtained in favorable cases by considering the reciprocal form
factor, $B({\mathbf r})$, which is defined as the Fourier transform of the
spin momentum density \cite{li04}:

\begin{equation}
B({\mathbf r})=\int \rho_{mag}({\mathbf p})
\exp(-i{\mathbf p}\cdot{\mathbf r}) d{\mathbf p}~.
\label{eq3}
\end{equation}
We now express $\rho_{mag}({\mathbf p})$ as a sum
over the momentum densities of the
molecular orbitals $\psi^{MO}_i({\mathbf p})$
of the magnetic electrons,
weighted by their occupancies $n_i$:
\begin{equation}
\rho_{mag}({\mathbf p})=\sum_i n_i |\psi^{MO}_i({\mathbf p})|^2~.
\label{eq4}
\end{equation}
By transforming Eq.~\ref{eq4} into real space, it is straightforwardly
shown that
\begin{equation}
B({\mathbf r})=\sum_{i}n_i\int \psi_i^{MO}({\mathbf s})
\psi_i^{*MO}({\mathbf s}+{\mathbf r}) d{\mathbf s},
\label{eq5}
\end{equation}
where the integrals on the right hand side give the autocorrelation of the
magnetic orbitals, $\psi^{MO}_i({\mathbf r})$, i.e. the overlap between
$\psi^{MO}_i({\mathbf s})$ and the same orbital translated by a distance
$\bf r$. Alternatively, $B({\mathbf r})$ along a given direction can be
obtained directly from Eq.~\ref{eq3} by taking the 1D-Fourier transform of
the
MCP along that direction. By modeling the right
hand side of Eq.~\ref{eq5}, and
comparing the results to the experimentally determined $B(r)$,
insight into
the occupation numbers $n_i$ can then be obtained.

\section{results and discussion}

\label{sec:res}
\begin{figure}
\begin{center}
\includegraphics[width=\hsize]{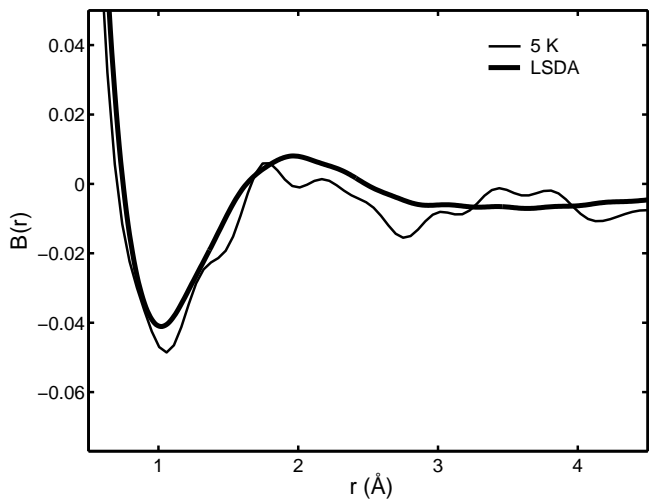}
\end{center}
\caption{
$B(r)$ as a function of $r$ along [110] obtained via a one-dimensional version
of the Fourier transform (Eq.~\ref{eq3}) of the experimental [110] MCP measured
at 5K (thin line), and the corresponding computed MCP (thick line) after
including the effect of experimental resolution broadening.
}
\label{fig1}
\end{figure}
\begin{figure}
\begin{center}
\includegraphics[width=\hsize]{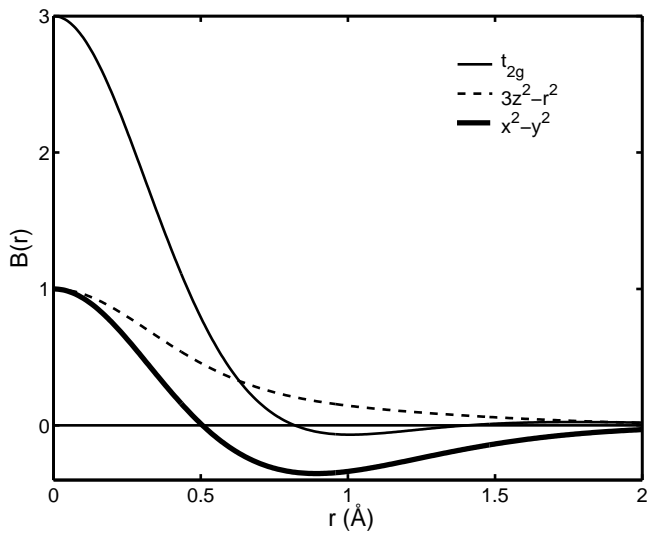}
\end{center}
\caption{
Partial contributions $B_{t2g}$, $B_{x^2-y^2}$ and $B_{3z^2-r^2}$ of
Eqs.~\ref{eq7}-\ref{eq9}
as a function of $r$. Note $B_{t2g}$ gives the total
contribution from the three $t_{2g}$
orbitals of $xy$, $yz$, and $zx$ symmetry. Computations are for $Z_{eff}=1$.
}
\label{fig2}
\end{figure}

Figure~\ref{fig1} compares the $B(r)$ along [110] obtained by using
Eq.~\ref{eq3}
from the measured MCP with the corresponding result from the computed MCP
based on the LSDA.\cite{footnote1} Our focus is on understanding the
nature of the
pronounced dip around $r \approx 1$ $\mbox{\AA}$,
which is seen in both theory and
experiment. In order to interpret this minimum in terms of occupancy of
magnetic orbitals, we model $B({\mathbf r})$ using Slater type orbitals
(STOs).\cite{weyrich,barnett} The STO of $xy$ symmetry for example is given
by
\begin{equation}
\psi_{xy}=N~xy \exp(-\xi r)~,
\label{Eq_sto}
\end{equation}
where $\xi$ is the exponent and $N$ is the normalization factor.
Other STOs are of similar form with the factor of $xy$ in Eq.~\ref{Eq_sto}
replaced by one of a different symmetry. The overlap integral in 
Eq.~\ref{eq5} for
various STOs is easily computed. The analytic form of the result is:
\begin{equation}
B_{t2g}=e^{-t}
\left (3+3\,t+t^{2}-\frac{11t^{4}}{140}-
\frac{t^{5}}{84}+\frac{t^{6}}{420}\right ),
\label{eq7}
\end{equation}
\begin{equation}
B_{x^2-y^2}=e^{-t}\left(1+t+\frac{2t^{2}}{7} -\frac{t^{3}}{21}-
\frac{t^{4}}{21}-\frac {t^{5}}{105}\right ),
\label{eq8}
\end{equation}
\begin{eqnarray}
\nonumber
B_{3z^2-r^2}&=&e^{-t}\left (1+t+\frac {8t^{2}}{21}
+\frac{t^{3}}{21} \right.\\
&&\left. -\frac{t^{4}}{140}
-\frac {t^{5}}{1260}+\frac {t^{6}}{1260}\right) ~,
\label{eq9}
\end{eqnarray}
where
\begin{equation}
t=\xi r =(Z_{eff}/3) r,
\label{Eq_Z}
\end{equation}
and $Z_{eff}$ denotes the effective nuclear charge
(all the quantities are implicitly in atomic units).
Note that we have
grouped the contributions of the three different $t_{2g}$ orbitals ($xy$,
$yz$, and
$zx$) into a single quantity $B_{t2g}$ in Eq.~\ref{eq7} for convenience. By
choosing an
occupancy of $3$ for the $t_{2g}$ states, of $f$ for the $d_{x^2-y^2}$ and
$(0.6-f)$ for the $d_{3z^2-r^2}$ state, the total $B(r)$ is given by
\begin{equation}
B(r)=\frac{B_{t2g}(r)+f B_{x^2-y^2}(r)+(0.6-f) B_{3z^2-r^2}(r)}{3.6}~,
\label{Eq_bmod}
\end{equation}
where we have imposed the condition $B(0)=1$ since we compare MCPs
normalized to one.

\begin{figure}
\begin{center}
\includegraphics[width=\hsize]{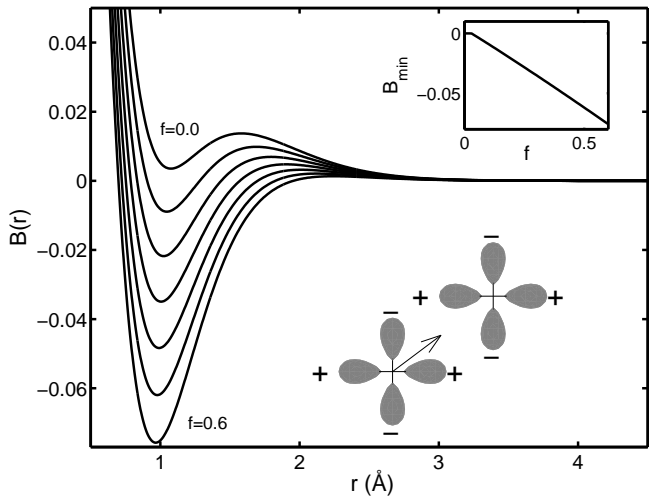}
\end{center}
\caption{
Model $B(r)$ function computed using Eq.~\ref{Eq_bmod} is shown for various
values of
the parameter $f$, which gives the occupancy of the $d_{x^2-y^2}$ orbital,
over
the range $f=0.0$ - $0.6$ (from top to bottom). Inset: Plot of minimum
$B_{min}$ of
$B(r)$ as a function of $f$ (see Eq.~\ref{Eq_bfit}). Diagram at bottom right:
A schematic
$d_{x^2-y^2}$ orbital centered at the origin with lobes pointed along [100]
and
[010] and the same orbital displaced along [110].
}
\label{fig3}
\end{figure}

Figures~\ref{fig2} and \ref{fig3} provide insight
into the modeling of $B(r)$ based on Eqs.~6-11.
Figure~\ref{fig2} shows the behavior of individual
contributions $B_{t2g}$,
$B_{x^2-y^2}$, and $B_{3z^2-r^2}$,
while Fig.~\ref{fig3} delineates the variations in the dip
around $r\approx 1$ $\mbox{\AA}$  as a function
of the occupancy $f$ of the $d_{x^2-y^2}$
orbital. We have chosen $Z_{eff}=1$ in all calculations in this article. It
is
clear from Eq.~\ref{Eq_Z} that $Z_{eff}$ is a scaling parameter. Its
precise value is
not so important for our analysis since it merely serves to shift the
position of the minimum in $B(r)$, but does not affect the size of the dip
in $B(r)$ which is our main concern. The key result of our analysis is that
the dip in $B(r)$ in Fig.~\ref{fig3}, which arises
from the contributions in $t^3$ and $t^4$,
is produced mainly by the autocorrelation of the $d_{x^2-y^2}$ orbital.
The $d_{3z^2-r^2}$ orbital is seen from Fig.~\ref{fig2} to give a positive
contribution near $1$ $\mbox{\AA}$.
With regard to the individual terms from the $xy$,
$yz$ and $zx$ orbitals to $B_{t2g}$
(not shown in Fig.~\ref{fig2}), the $d_{xy}$ orbital
gives a small positive contribution,
while the $d_{xz}$ and $d_{yz}$ orbitals
give negative contributions.

The minimum value $B_{min}$ of $B(r)$ along [110] is seen from
Fig.~\ref{fig3} to
be correlated with the $d_{x^2-y^2}$ occupancy given by $f$. The inset to
Fig.~\ref{fig3} in fact shows that $B_{min}$ can be fitted linearly:
\begin{equation}
B_{min}=af+b,
\label{Eq_bfit}
\end{equation}
where $a=-0.130$ and $b=0.004$. The sketch of the $d_{x^2-y^2}$ orbital in
Fig.~\ref{fig3} is perhaps helpful in this regard: when the $d_{x^2-y^2}$
orbital
is translated along [110], the positive and negative lobes will overlap
and yield a {\em negative} dip at a distance of the order of orbital
dimensions. Finally, we note that in writing Eq.~\ref{eq3} for $B(r)$, we
have
not included the effect of finite experimental resolution on the MCP.
The finite resolution $\Delta p$ (FWHM) in momentum space (assuming a
gaussian resolution function) leads to a multiplicative
attenuation of $B(r)$ in $r$-space given by
\begin{equation}
B_a(r)=B(r)\exp(-r^2/S^2),
\end{equation}
where
\begin{equation}
S=\frac{4\sqrt{\ln(2)}}{\Delta p}.
\end{equation}
For $\Delta p=0.4$ a.u.
the exponential factor is 95\% at
$r= 1~\mbox{\AA}$, therefore it has
little effect on the value of $B_{min}$.

The minimum in the $B(r)$ curves of Fig.~\ref{fig1} can now be used to
derive the
occupancies of the $e_g$ orbitals.
Using the fit of Eq.~\ref{Eq_bfit}, we thus obtain
for the LSDA curve of Fig.~\ref{fig1}, $f=0.35$, which implies occupancies
of
$0.35$ and $0.25$ for the $d_{x^2-y^2}$ and $d_{3z^2-r^2}$ orbitals,
respectively. The corresponding values for the 5K experimental $B(r)$ of
Fig.~\ref{fig1} are: $0.40$ for $d_{x^2-y^2}$ and $0.20$ for
$d_{3z^2-r^2}$.\cite{footnote2}
This preferential occupation of the $d_{x^2-y^2}$ orbital is somewhat
surprising since the Mn-O bond length for the apical O atoms is larger
than that for the in-plane O atoms, and a simple molecular orbital scheme
would suggest that the energy of the antibonding $d_{3z^2-r^2}$ level will
be lowered in relation to that of the $d_{x^2-y^2}$ level. Note, however,
that electron correlations beyond the LSDA are needed to stabilize the
canted antiferromagnetic order between the two Mn-O planes observed in the
absence of an external magnetic field. The occupancy of the $d_{3z^2-r^2}$
orbital promotes ferromagnetism,\cite{koizumi01} so that
antiferromagnetic correlations will tend to reduce the $d_{3z^2-r^2}$
population. Reference \onlinecite{med01} argues that the on-site Coulomb
correlation $U$ of Mn $d$ electrons can produce a significant splitting of
the $e_g$ states, modify interlayer exchange interactions and magnetic
ordering, and promote anisotropy in the electrical transport by reducing
the conduction along the $c$-axis. Bearing all this in mind, we consider
the discrepancy between the depth of the minimum between the LSDA and
experiment in Fig.~\ref{fig1} to reflect the presence of electron
correlation effects beyond the LSDA framework underlying our
computations.\cite{maekawa01,oles03}

Turning to the behavior of $B(r)$ beyond the dip around $r \approx 1$ 
$\mbox{\AA}$, Fig.~\ref{fig1} shows the presence of long range 
oscillations in the experimental $B(r)$ as well as the corresponding LSDA 
calculations (damping effect of the resolution notwithstanding). These 
oscillations at large $r$ are absent in our relatively simple STO-based 
model as seen from Fig. 3. We emphasize that sizable values of $B(r)$ 
beyond atomic dimensions are a hallmark of electronic states extending 
over larger distances as a result of the mixing of Mn and O orbitals in 
the MnO$_2$ planes. These hybridization effects are crucially important 
for faithfully modeling the shape of the MCP in a global 
sense.\cite{koizumi01,koizumi03} Our relatively simple STO-based model is 
not meant to reproduce such a global fit to the MCP, but to only highlight 
the atomic-like features hidden in the MCP via the use of the $B(r)$ 
function at low $r$-values. As we have already pointed out, when the x-ray 
scattering vector lies along the [110] direction, the number of magnetic 
electrons of a {\em specific} symmetry, i.e., $d$-electrons of $x^2-y^2$ 
symmetry, yield a distinct atomic signature in the $B(r)$, allowing us to 
extract the occupancies of the $e_g$ electrons.

\section{summary and conclusions}
\label{sec:sum}

We discuss how the shape of the [110] MCP in
La$_{1.2}$Sr$_{1.8}$Mn$_2$O$_7$
can be analyzed in
terms of the reciprocal form factor $B(r)$ given by the one-dimensional
Fourier transform of the MCP. By interpreting $B(r)$ as the autocorrelation
function of the ground state wavefunction, one can obtain a handle on the
occupancies of various magnetic orbitals in the system. In particular, the
$B(r)$ along the [110] direction in La$_{1.2}$Sr$_{1.8}$Mn$_2$O$_7$
contains a distinct dip at $r\approx 1$ $\mbox{\AA}$
and the value $B_{min}$ of $B(r)$ at the minimum can be related directly
to the occupancy $f$ of the $d_{x^2-y^2}$ orbital as:
$B_{min}=af+b$, where $a=-0.130$ and $b=0.004$.
The specific forms of the $B(r)$ functions
associated with various $t_{2g}$ and $e_g$ Slater-type orbitals are
presented
and used to analyze $B(r)$ functions for the 5K MCP data and the computed
MCP based on the LSDA as illustrative examples. In this way, for LSDA we
adduce
$f=0.35$ or occupancies of $0.35$ for $d_{x^2-y^2}$ and $0.25$
for $d_{3z^2-r^2}$ orbitals; the
corresponding occupancies derived from the experimental 5K MCP are $0.40$
for
$d_{x^2-y^2}$ and $0.20$ for the $d_{3z^2-r^2}$ orbital.
The larger experimental value of the
$d_{x^2-y^2}$ occupancy reflects limitations of the LSDA in accounting for
electron correlation effects in the system. Our approach would allow a
determination of the occupancies of magnetic orbitals through an analysis
of the shapes of the MCPs more generally in complex materials.

\begin{acknowledgments}

We acknowledge discussions with Robert Markiewicz.
This work is supported by the US Department of Energy under contract
DE-AC03-76SF00098. It is also sponsored by the Stichting Nationale Computer
Faciliteiten (NCF) for the use of supercomputer facilities, with financial
support from the Netherlands Organization for Scientific Research (NWO),
and benefited from the allocation of supercomputer time at the NERSC and
the Northeastern University's Advanced Scientific Computation Center
(NU-ASCC).

\end{acknowledgments}


\begin{thebibliography}{99}
\bibitem{footnote0.5}
  In our computations the core states are
  treated relativistically, but the valence electrons are treated
  semi-relativistically.
\bibitem{footnote2}

Values of occupancies $f$ in Ref.~\onlinecite{li04} were obtained by neglecting the 
effect of the small offset given by the parameter $b=0.004$ in Eq.~\ref{Eq_bfit}
and differ slightly from those given here. Notably, we expect the error bars 
on $B(r)$ (and thus on $B_{min}$) to be quite small since $B(r)$ involves 
an integral over the MCP.
\end{thebibliography}
\end{document}